\documentclass[12pt]{article}
\usepackage{amsfonts}
\usepackage[ruled,vlined]{algorithm2e}
\usepackage{amsmath,amssymb}
\usepackage{graphicx}

\newlength{\pagewidth}
\newcommand{\commentOut}[1]{}
\newcommand{\derives}[1][*]{\stackrel{#1}{\Rightarrow}}

\title{An error correcting parser for context free grammars\\ that takes less than cubic time}
\author{Sanguthevar Rajasekaran and Marius Nicolae\\
Department of CSE, Univ. of Connecticut, Storrs}
 
\newtheorem{theorem}{Theorem}[section]

\begin{document}
\date{}
\maketitle

\abstract{The problem of parsing has been studied extensively for various formal grammars. Given an input string and a grammar, the parsing problem is to check if the input string belongs to the language generated by the grammar. A closely related problem of great importance is one where the input are a string ${\cal I}$ and a grammar $G$ and the task is to produce a string ${\cal I}'$ that belongs to the language generated by $G$ and the `distance' between ${\cal I}$ and ${\cal I}'$ is the smallest (from among all the strings in the language). Specifically, if ${\cal I}$ is in the language generated by $G$, then the output should be ${\cal I}$. Any parser that solves this version of the problem is called an {\em error correcting parser}. In 1972 Aho and Peterson presented a cubic time error correcting parser for context free grammars. Since then this asymptotic time bound has not been improved under the (standard) assumption that the grammar size is a constant. In this paper we present an error correcting parser for context free grammars that runs in $O(T(n))$ time, where $n$ is the length of the input string and $T(n)$ is the time needed to compute the tropical product of two $n\times n$  matrices.}

In this paper we also present an $\frac{n}{M}$-approximation algorithm for the {\em language edit distance problem} that has a run time of $O(Mn^\omega)$, where $O(n^\omega)$ is the time taken to multiply two $n\times n$ matrices. To the best of our knowledge, no approximation algorithms have been proposed for error correcting parsing for general context free grammars.

\section{Introduction}
Parsing is a well studied problem owing to its numerous applications. For example, parsing finds a place in programming language translations, description of properties of semistructured data \cite{Sah13}, protein structures prediction \cite{RAA10}, etc. For context free grammars, two classical algorithms can be found in the literature: CYK \cite{CoSc70,You67,Kas65} and Earley \cite{Ear70}. Both of these algorithms take $O(n^3)$ time in the worst case. Valiant has shown that context free recognition can be reduced to Boolean matrix multiplication \cite{Val75}. 

The problem of parsing with error correction (also known as the {\em language edit distance problem}) has also been studied well. Aho and Peterson presented an $O(n^3)$ time algorithm for context free grammar parsing with errors. Three kinds of errors were considered, namely, insertion, deletion, and substitution. This algorithm depended quadratically on the size of the grammar and was based on Earley parser. Subsequently, Myers \cite{Mye95} presented an algorithm for error correcting parsing for context free grammars that also runs in cubic time but the dependence on the grammar size was linear. This algorithm is based on the CYK parser.


As far as the worst case run time is concerned, to the best of our knowledge, cubic time is the best known for the error correcting parsing problem for general context free grammars. A number of approximation algorithms have been proposed for the DYCK language (which is a very specific context free language). See e.g., \cite{Sah13}.

In this paper we present a cubic time algorithm for error correcting parsing
 that is considerably simpler than the
algorithms of \cite{AhPe72} and \cite{Mye95}. This algorithm is based on the CYK
parser. Even though the algorithm of \cite{Mye95} is also based on CYK parser,
there are some crucial differences between our algorithm and that of
\cite{Mye95}. We also show that the language edit distance problem can be
reduced to the problem of computing the tropical product (also known as the
distance product or the min-plus product) of two given $n\times n$ matrices
where $n=|{\cal I}|$. Using the current best known run time \cite{Chan10} for
tropical matrix product, our reduction implies that the language edit distance
problem can be solved exactly in $O\left(\frac{n^3(\log\log n)^3}{(\log n)^2}\right)$ time,
improving the cubic run time that has remained the best since 1972.

In many applications, it may suffice to solve the language edit distance problem approximately. To the best of our knowledge, no approximation algorithms are known for general context free grammars. However, a number of such algorithms have been proposed for the Dyck language. Dyck language is a basic and fundamental context free grammar and the Dyck language edit distance is a significant generalization of string edit distance problem which has been widely studied. Also, the non-deterministic Dyck language is the hardest context free grammar in terms of parsing. The concept of approximate error correcting parsing was introduced by \cite{KSS+13}. 
The algorithm of \cite{KSS+13} takes subcubic time but its approximation factor is $\Theta(n)$. If ${\cal I}'$ is a string in the language generated by the input grammar $G$ that has the minimum language edit distance (say $d$) with the input string $\cal I$ and if an algorithm $\cal A$ outputs a string ${\cal I}''$ such that the language edit distance between $\cal I$ and ${\cal I}''$ is no more than $d\beta(n)$, then we say that $\cal A$ is a $\beta(n)$-approximation algorithm.
 Saha \cite{Sah13} gives the very first near-linear time algorithm for Dyck language edit distance problem with polylog approximaiton and an $O(n^{1+\epsilon}+n^\epsilon d^2)$ time algorithm with $1/\epsilon \log(d)$ approximation where $d$ is the edit
distance. \cite{Sah13} not only studies Dyck language edit distance problem, but also a larger class
of problems including the memory checking languages, which comprise of transcripts of any
popular data structure. It can also be applied to other variants of languages studied by Parnas, Ron and Rubinfeld (APPRO-RANDOM, 2001). It is noteworthy that Dyck grammar
parsing (without error correction) can easily be done in linear time. On the other hand, it
is known that parsing of arbitrary context free grammars is as difficult as boolean matrix
multiplication \cite{Lee02}. For an extensive discussion on approximation algorithms for the Dyck
language, please see \cite{Sah13}. In this paper we present an approximation algorithm for general context free grammars. Specifically, we show that if we are only interested in edit distances
of no more than $M$, then the language edit distance problem can be solved in $O((M+1)n^\omega)$
time where $O(n^\omega$) is the time taken to multiply two $n\times n$ matrices. (Currently the best
known value for $\omega$ is 2.376). As a corollary, it follows that there is an $\frac{n}{M}$-approximation
algorithm for the language edit distance problem with a run time of $O(Mn^\omega)$.

\subsection{Some Notations}
A context free grammar $G$ is a 4-tuple $(N, T, P, S)$, where $T$ is a set of characters (known as terminals) in the alphabet, $N$ is a set of variables known as nonterminals, $S$ is the start symbol (that is a nonterminal) and $P$ is a set of productions. 

We use $L(G)$ to denote the language generated by $G$. Capital letters such as $A,B,C,\ldots$ will be used to denote nonterminals, small letters such as $a,b,c,\ldots$ will be used to denote terminals, and greek letters such as $\alpha,\beta,\ldots$ will be used to denote any string from $(N\cup T)^*$.

A production of the form $A\rightarrow\epsilon$ is called an $\epsilon$-production. A production of the kind $A\rightarrow B$ is known as a unit production.

Let $G$ be a CFG such that $L(G)$ does not have $\epsilon$. Then we can convert $G$ into Chomsky Normal Form (CNF). A context free grammar is in CNF if the productions in $P$ are of only two kinds: $A\rightarrow a$ and $A\rightarrow BC$.

Let $U$ and $V$ be two $n\times n$ real matrices. Then the tropical (or distance) product $Z$ of $X$ and $Y$ is defined as: $Z_{ij}=\min_{k=1}^n (X_{ik}+Y_{kj}),~1\leq i,j\leq n$. 

The edit distance between two strings $\cal I$ and $\cal I'$ from an alphabet $T$ is the minimum number of (insert, delete, and substitution) operations needed to convert $\cal I$ to $\cal I'$.

In this paper we assume that the grammar size is $O(1)$ which is a standard assumption made in many works (see e.g., \cite{Val75}).

\subsection{Some Preliminaries}
\subsubsection{A summary of Aho and Peterson's algorithm}
The algorithm of Aho and Peterson \cite{AhPe72} is based on the parsing algorithm of Earley \cite{Ear70}. There are some crucial differences. Let ${\cal I}=a_1a_2\ldots a_n$ be the input string. If $G(N,T,P,S)$ is the input grammar, another grammar $G'=(N',T,P',S')$ is constructed where $G'$ has all the productions of $G$ and some additional productions that can be used to make derivations involving errors. Each such additional production is called an {\em error production}. Three kinds of errors are considered, namely, insertion, deletion, and substitution. $G'$ also has some additional nonterminals. The algorithm derives the input string beginning with $S'$, minimizing the number of applications of the error productions. 

The parser of \cite{AhPe72} can be thought of as a modified version of the
Earley parser. Like the algorithm of Earley, $n+1$ levels of lists are
constructed. Each list consists of items where an item is an object of the form
$[A\rightarrow\alpha.\beta,i,k]$. Here $A\rightarrow \alpha\beta$ is a
production, $.$ is a special symbol that indicates what part of the production
has been processed so far, $i$ is an integer indicating input position at which
the derivation of $\alpha$ started, and $k$ is an integer indicating the number
of error productions that have been used in the derivation from $\alpha$. If we
use ${\cal L}_j$ to denote the list of level $j,0\leq i\leq n$, then the item
$[A\rightarrow\alpha.\beta,i,k]$ will be in ${\cal L}_j$ if and only if for
some $\nu$ in $(N\cup T)^*$, $S'\derives a_1a_2\cdots a_iA\nu$ and
$\alpha\derives a_{i+1}a_{i+2}\cdots a_j$ using $k$ error productions.

The algorithm constructs the lists ${\cal L}_0,{\cal L}_1,\ldots,{\cal L}_n$. An item of the form $[S'\rightarrow \alpha.,0,k]$ will be in ${\cal L}_n$, for some integer $k$. In this case, $k$ is the minimum edit distance between ${\cal I}$ and any string in $L(G)$.

Note that the Earley parser also works in the same manner except that an item will only have two elements: $[A\rightarrow\alpha.\beta,i]$.

\subsubsection{A synopsis of Valiant's algorithm}\label{Valiant}
Valiant has presented an efficient algorithm for computing the transitive closure of an
upper triangular matrix. The transitive closure
is with respect to matrix multiplication defined in a specific way. Each
element in a matrix will be a set of items. In the case of context free
recognition, each matrix element will be a set of nonterminals. If $N_1$ and
$N_2$ are two sets of nonterminals, a binary operator $\cdot$ is defined as: 
$N_1 \cdot N_2=\{A|\exists B\in N_1,C\in N_2$ such that $(A\rightarrow BC)\in
P\}$. If $a$ and $b$ are matrices where each element is a subset of $N$, the
product matrix $c$ of $a$ and $b$ is defined as follows.
$$c_{ij}=\bigcup_{k=1}^na_{ik} \cdot b_{kj}$$

Under the above definition of matrix multiplication, we can define transitive closure for any matrix $a$ as:
$$a^+=a^{(1)}\cup a^{(2)}\cup\cdots$$
where $a^{(1)}=a$ and
$$a^{(i)}=\bigcup_{j=1}^{i-1}a^{(j)} \cdot a^{(i-j)}$$

Valiant has shown that this transitive closure can be computed in $O(S(n))$
time, where $S(n)$ is the time needed for multiplying two matrices with the
above special definition of matrix product. In fact this algorithm works for
the computation of transitive closure for generic operators $\odot$ and union
as long as these operations satisfy the following properties: The outer
operation (i.e., union) is commutative and associative, the inner operation
($\odot$) distributes over union, $\emptyset$ is a multiplicative zero and an
additive identity.

\section{A Simple Error Correcting Parser}
In this section we present a simple error correcting parser for CFGs. This algorithm is based on the algorithm of \cite{Raj96}. We also utilize the concept of error productions introduced in \cite{AhPe72}. If $G=(N,T,P,S)$ is the input grammar, we generate another grammar $G'=(N',T,P',S)$ where $N'=N\cup\{H,I\}$. $P'$ has all the productions in $P$. In addition, $P'$ has some additional error productions. We parse the given input string ${\cal I}$ using the productions in $P'$. For each production, we associate an error count that indicates the minimum number of errors the use of the production will amount to. The goal is to parse ${\cal I}$ using as few error productions as possible. Specifically, the sum of error counts of all the error productions used should be minimum.
If $A\rightarrow \alpha$ is an error production with an error count of $k$, we denote this rule as $A\stackrel{k}{\rightarrow}\alpha$. If there is no integer above $\rightarrow$ in any production, the error count of this production should be assumed to be $0$.

\subsection{Construction of a covering grammar}\label{ggen}
Let $G=(N,T,P,S)$ be the given grammar and ${\cal I}=a_1a_2\ldots a_n$ be the given input string. Without loss of generality assume that $L(G)$ does not have $\epsilon$ and that $G$ is in CNF. Even if $G$ is not in CNF, we could employ standard techniques to convert $G$ into this form (see e.g., \cite{HMU06}). We construct a new grammar $G'=(N',T,P',S)$ as follows. $P'$ has the following productions in addition to the ones in $P$:  $H\stackrel{0}{\rightarrow} HI$, $H\stackrel{0}{\rightarrow} I$, and $I\stackrel{1}{\rightarrow} a$ for every $a\in T$. Here $H$ and $I$ are new nonterminals. If $A\rightarrow a$ is in $P$, then add the following rules to $P'$: $A\stackrel{1}{\rightarrow} b$ for every $b\in T-\{a\}$, $A\stackrel{1}{\rightarrow}\epsilon$, $A\stackrel{0}{\rightarrow} AH$, and $A\stackrel{0}{\rightarrow} HA$. Each production in $P$ has an error count of $0$.

\vspace{0.2in}

\noindent{\bf Elimination of $\epsilon$-productions:} We first eliminate the $\epsilon$-productions in $P'$ as follows. We say a nonterminal $A$ is nullable if $A\derives \epsilon$. Let $k$ be the number of errors needed for $A$ to derive $\epsilon$. We denote this as follows: $A\stackrel{*,k}{\Rightarrow}\epsilon$. Call $k$ the $nullcount$ of $A$, denoted as $nullcount(A)$. We only keep the minimum such $nullcount$ for any nonterminal. Let the minimum $nullcount$ for any terminal $A$ be $Mnullcount(A)$. For example, if $A\stackrel{0}{\rightarrow}BC,B\stackrel{1}{\rightarrow}\epsilon$,
and $C\stackrel{1}{\rightarrow}\epsilon$ are in $P'$, then $A\stackrel{*,2}{\Rightarrow}\epsilon$. We identify all the nullable nonterminals in $P'$ using the following procedure. If $B\rightarrow CD$ is in $P'$ and if both $C$ and $D$ are nullable, then $B$ is nullable as well. In this case, $nullcount(B)=nullcount(C)+nullcount(D)$.

After identifying all nullable nonterminals and their $Mnullcount$ values, we process each production as follows. Let $A\stackrel{k}{\rightarrow} BC$ be any production in $P'$. If $B$ is nullable and $C$ is not, and if $Mnullcount(B)=\ell$, then we add the production $A\stackrel{k+\ell}{\rightarrow}C$ to $P'$. If $C$ is nullable and $B$ is not, and if $Mnullcount(C)=\ell$, then we add the production $A\stackrel{k+\ell}{\rightarrow}B$ to $P'$. If both or none of $B$ and $C$ are nullable, then we do not add any additional production to $P'$ while processing the production $A\stackrel{k}{\rightarrow} BC$. If there are more than one productions in $P'$ with the same precedent and consequent, we only keep that production for which the error count is the least. 

Finally, we remove all the $\epsilon$ productions.

\vspace{0.2in}

\noindent{\bf Elimination of unit productions:} We eliminate unit productions from $P'$ as follows. Let $A\stackrel{k_1}{\rightarrow}B_1,B_1\stackrel{k_2}{\rightarrow}B_2,\ldots,B_{q-3}\stackrel{k_{q-2}}{\rightarrow}B_{q-2},B_{q-2}\stackrel{k_{q-1}}{\rightarrow}B$ be a sequence of unit productions in $P'$ and $B\stackrel{k_q}{\rightarrow}\alpha$ be a non unit production. In this case we add the production $A\stackrel{Q}{\rightarrow}\alpha$ to $P'$, where $Q=\sum_{i=1}^qk_i$. After processing all such sequences and adding productions to $P'$ we eliminate duplicates. In particular, if there are more than one rules with the same precedent and consequent, we only keep the production with the least error count. At the end we remove all the unit productions.

\vspace{0.2in}

\noindent {\bf Observation:} Aho and Peterson \cite{AhPe72} indicate that $G'$ is a covering grammar for $G$ and prove several properties of $G'$. Note that they don't keep any error counts with their productions. Also, the validity of the procedures we have used to eliminate $\epsilon$ and unit productions can be found in \cite{HMU06}.

\vspace{0.2in}

\noindent{\bf An Example.} Consider the language $\{a^nb^n:n\geq 1\}$. A CFG for this language has the productions: $S\rightarrow aSb|ab$. We can get an equivalent grammar $G=(N,T,P,S)$ in CNF where $N=\{S,A,B,A_1\},T=\{a,b\},$ and $P=\{S\rightarrow AA_1|AB, A_1\rightarrow SB,A\rightarrow a, B\rightarrow b\}$.

We can get a grammar $G_1=(N',T,P_1,S)$ with error productions where $N'=\{S,A,B$, $A_1,H,I\}$ and $P_1=\{S\rightarrow AA_1|AB, A_1\rightarrow SB,A\rightarrow a, B\rightarrow b, H\rightarrow HI, H\rightarrow I, I\stackrel{1}{\rightarrow}a, I\stackrel{1}{\rightarrow}b, A\stackrel{1}{\rightarrow}b,A\stackrel{1}{\rightarrow}\epsilon,A\rightarrow HA,A\rightarrow AH,B\stackrel{1}{\rightarrow}a,B\stackrel{1}{\rightarrow}\epsilon,B\rightarrow HB,B\rightarrow BH\}$. Note that any production with no integer above $\rightarrow$ has an error count of zero.

\begin{itemize}
\item{\bf Eliminating $\epsilon$-productions:} We identify nullable
nonterminals. We realize that the following nonterminals are nullable:
$A,B,S$, and $A_1$. For example, $A_1$ is nullable since we have:
$A_1\stackrel{*}{\Rightarrow} SB\stackrel{*}{\Rightarrow} ABB\stackrel{*,1}{\Rightarrow}
BB\stackrel{*,1}{\Rightarrow} B\stackrel{*,1}{\Rightarrow} \epsilon$.
We also realize: $A\stackrel{*,1}{\Rightarrow}\epsilon,B\stackrel{*,1}{\Rightarrow}\epsilon,S\stackrel{*,2}{\Rightarrow}\epsilon$, and $A_1\stackrel{*,3}{\Rightarrow}\epsilon$.

Now we process every production in $P_1$ and generate new relevant rules. For instance, consider the rule $S\rightarrow AA_1$. Since $A_1\stackrel{*,3}{\Rightarrow}\epsilon$, we add the rule $S\stackrel{3}{\rightarrow}A$ to $P_1$. When we process the rule $S\rightarrow AB$, since $B\stackrel{*,1}{\Rightarrow}\epsilon$, we realize that $S\stackrel{1}{\rightarrow}A$ has to be added to $P_1$. However, $S\stackrel{3}{\rightarrow}A$ has already been added to $P_1$. Thus we replace $S\stackrel{3}{\rightarrow}A$ with $S\stackrel{1}{\rightarrow}A$. 

Processing in a similar manner, we add the following productions to $P_1$ to get $P_2$: $S\stackrel{1}{\rightarrow}A_1,S\stackrel{1}{\rightarrow}A,S\stackrel{1}{\rightarrow}B, A_1\stackrel{1}{\rightarrow}S, A_1\stackrel{2}{\rightarrow}B$, $A\stackrel{1}{\rightarrow}H$, and $B\stackrel{1}{\rightarrow}H$. We eliminate all the $\epsilon$-productions from $P_2$.

\item {\bf Eliminating unit productions:} We consider every sequence of unit productions $A\stackrel{k_1}{\rightarrow}B_1,B_1\stackrel{k_2}{\rightarrow}B_2,\ldots,B_{q-3}\stackrel{k_{q-2}}{\rightarrow}B_{q-2},B_{q-2}\stackrel{k_{q-1}}{\rightarrow}B$ in $P_2$ with $B\stackrel{k_q}{\rightarrow}\alpha$ being a non unit production. In this case we add the production $A\stackrel{Q}{\rightarrow}\alpha$ to $P_2$, where $Q=\sum_{i=1}^qk_i$.

Consider the sequence $S\stackrel{1}{\rightarrow} A,A\rightarrow a$. This sequence results in a new production: $S\stackrel{1}{\rightarrow} a$. The sequence $S\stackrel{1}{\rightarrow} A_1,A_1\stackrel{2}{\rightarrow}B,B\stackrel{1}{\rightarrow}a$ suggests the addition of the production $S\stackrel{4}{\rightarrow}a$. But we have already added a better production and hence this production is ignored. 

Proceeding in a similar manner we realize that we have to add the following productions to $P_2$ to get $P_3$: $S\stackrel{1}{\rightarrow}a,S\stackrel{1}{\rightarrow}b,H\stackrel{1}{\rightarrow}a,H\stackrel{1}{\rightarrow}b,A_1\stackrel{2}{\rightarrow}a$, $A_1\stackrel{2}{\rightarrow}b$, 
$S\stackrel{1}{\rightarrow}HB$, $S\stackrel{1}{\rightarrow}BH$,
$A\stackrel{1}{\rightarrow}HI$, $B\stackrel{1}{\rightarrow}HI$, $S\stackrel{2}{\rightarrow}HI$, $A_1\stackrel{3}{\rightarrow}HI$, $A_1\stackrel{2}{\rightarrow}BH$, $A_1\stackrel{2}{\rightarrow}HB$, $S\stackrel{1}{\rightarrow}AH$, $S\stackrel{1}{\rightarrow}HA$, $A_1\stackrel{2}{\rightarrow}AH$, and $A_1\stackrel{2}{\rightarrow}HA$ .  We eliminate all the unit productions from $P_3$.
\end{itemize}

The final grammar we get is $G_3=(N',T,P_3,S)$ where $P_3=\{S\rightarrow AA_1|AB, A_1\rightarrow SB,A\rightarrow a, B\rightarrow b, H\rightarrow HI, I\stackrel{1}{\rightarrow}a, I\stackrel{1}{\rightarrow}b, A\stackrel{1}{\rightarrow}b,A\rightarrow HA,A\rightarrow AH,B\stackrel{1}{\rightarrow}a,B\rightarrow HB,B\rightarrow BH, S\stackrel{1}{\rightarrow}a,S\stackrel{1}{\rightarrow}b,H\stackrel{1}{\rightarrow}a,H\stackrel{1}{\rightarrow}b,A_1\stackrel{2}{\rightarrow}a, A_1\stackrel{2}{\rightarrow}b, 
 S\stackrel{1}{\rightarrow}HB, S\stackrel{1}{\rightarrow}BH,
A\stackrel{1}{\rightarrow}HI, B\stackrel{1}{\rightarrow}HI, S\stackrel{2}{\rightarrow}HI, A_1\stackrel{3}{\rightarrow}HI, A_1\stackrel{2}{\rightarrow}BH, A_1\stackrel{2}{\rightarrow}HB, S\stackrel{1}{\rightarrow}AH, S\stackrel{1}{\rightarrow}HA, A_1\stackrel{2}{\rightarrow}AH, A_1\stackrel{2}{\rightarrow}HA \}$.

\subsection{The algorithm}\label{simple}
The algorithm is a modified version of an algorithm given in \cite{Raj96}.
This algorithm in turn is a slightly different version of the CYK algorithm. 
Let $G' =(N', T', P', S')$ be the grammar generated using the procedure given in Section~\ref{ggen}.
The basic idea behind the algorithm is the following: The algorithm has $n$ stages.
In any given stage we scan through each production in $P'$ and grow larger and larger parse
trees. At any given time in the algorithm, each nonterminal has a list of tuples of the form $(i, j,\ell)$. If $A$
is any nonterminal, $LIST(A)$ will have tuples $(i, j,\ell)$ such that $A\stackrel{*,\ell}{\Rightarrow}a_{i}\ldots a_{j-1}$ and there is no $\ell'<\ell$ such that  $A\stackrel{*,\ell'}{\Rightarrow}a_{i}\ldots a_{j-1}$.
If $A\stackrel{\ell_3}{\rightarrow} BC$ is a production in $P'$, then in any stage we process this production as follows:
We scan through elements in $LIST(B)$ and look for matches in $LIST(C)$. For instance, if
$(i, k,\ell_1)$ is in $LIST(B)$ (for some integer $\ell_1$), we check if $(k,
j,\ell_2)$ is in $LIST (C)$, for some $j$ and $\ell_2$. If so, we insert $(i,
j,\ell_1+\ell_2+\ell_3)$ into $LIST(A)$. If $LIST(A)$ for any $A$ has many
tuples of the form $(i,j,*)$ we keep only one among these. Specifically, if
$(i,j,\ell_1),(i,j,\ell_2),\ldots,(i,j,\ell_q)$ are in $LIST(A)$, we keep only
$(i,j,\ell)$ where $\ell=\min_{m=1}^q\ell_m$.

 We maintain the following data structures:  (1) for each nonterminal $A$, an array (call
it $X_A$) of lists indexed 1 through $n$, where $X_A[i]$ is the list of all tuples from $LIST(A)$ whose
first item is $i~(1 \leq i\leq n)$; and (2) an $n\times n$ upper triangular matrix $\cal M$ whose $(i, j)$th entry will be those
nonterminals that derive $a_{i}a_{i+1}\ldots a_{j-1}$ with the corresponding (minimum) error counts (for $1\leq i\leq n$ and $2\leq j\leq(n+1)$). There can be $O(n^2)$ entries in $LIST(B)$, and for
each entry $(i, k)$ in this list, we need to search for at most $n$ items in $LIST (C)$.

By induction, we can show that at the end of stage $s~(1 \leq s\leq n$), the algorithm would
have computed all the nonterminals that span any input segment of length $s$ or less, together with the minimum error counts. (We say
a nonterminal spans the input segment $J= a_{i}a_{i+1}\ldots a_{j-1}$ if it derives $J$; the nonterminal
is said to have a "span-length" of $j-i$.)

\begin{figure}[h]
\centering
\includegraphics[width=\textwidth]{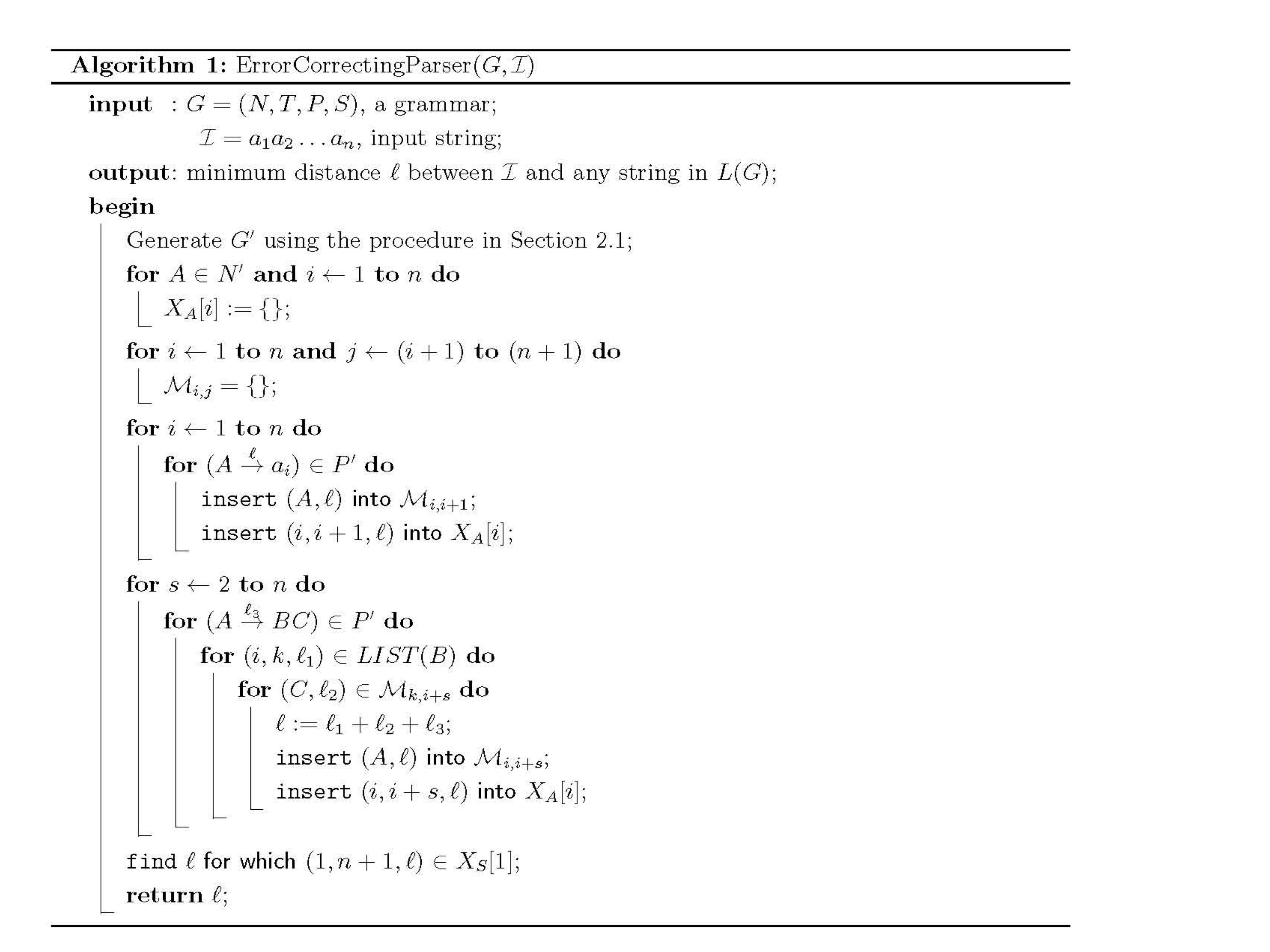}
\vspace*{-0.2in}
\caption{\label{fig1} Algorithm 1.}
\end{figure}

A straight forward implementation of the above idea takes $O(n^4)$ time. However, we can reduce the run time of each stage to $O(n^2)$ as follows: In stage $s$, while
processing the production $A \rightarrow BC$, work only with tuples from $LIST(B)$ and $LIST(C)$
whose combination will derive an input segment of length exactly $s$. For example, if $(i, k,\ell)$ is
a tuple in $LIST(B)$, the only tuple in $C$ we should look for is $(k, i+s,\ell')$ (for any integer $\ell'$). We can look for
such a tuple in $O (1)$ time using the matrix $\cal M$. With this modification, each stage of the above
algorithm will only take $O (n^2)$ time and hence the total run time of the
algorithm is $O(n^3)$. The pseudocode is given in algorithm \ref{fig1}.

\begin{theorem}
When the above algorithm completes, for any nonterminal $A$, $LIST(A)$ has a tuple $(i,j,\ell)$ if and only if $A\stackrel{*,\ell}{\Rightarrow}a_{i}\ldots a_{j-1}$ and there is no $\ell'<\ell$ such that $A\stackrel{*,\ell'}{\Rightarrow}a_{i}\ldots a_{j-1}$.
\end{theorem}

\noindent{\bf Proof:} The proof is by induction on the stage number $s$.

\noindent{\bf Base Case:}  is when $s=1$, i.e., $j-i=1$, for $1\leq i,j\leq (n+1)$. Note that all the nonterminals other than $H$ and $I$ are nullable. As a result, $P'$ will have a production of the kind $A\stackrel{\ell}{\rightarrow} b$ for every nonterminal $A$ and every terminal $b$, for some integer $\ell$. By the way we compute $Mnullcount$ for each nonterminal and eliminate unit productions, it is clear that $\ell$ is the smallest integer for which $A\stackrel{*,\ell}{\Rightarrow}b$.

\vspace{0.1in}

\noindent{\bf Induction Step:} Assume that the hypothesis is true for span lengths up to $s-1$. We can prove it for a span length of $s$. Let $A\stackrel{l_3}{\rightarrow}BC$ be any production in $P'$. Let $(i,k,\ell_1)$ be a tuple in $LIST(B)$ and $(k,j,\ell_2)$ be a tuple in $LIST(C)$ with $j-i=s$. Then this means that $A\stackrel{*,L}{\Rightarrow}a_{i}\ldots a_{j-1}$, where $L=\ell_1+\ell_2+\ell_2$. We add the tuple $(i,j,L)$ to $LIST(A)$. Also, the induction hypothesis implies that $B\stackrel{*,\ell_1}{\Rightarrow}a_{i}\ldots a_{k-1}$ and there is no $\ell<\ell_1$ for which $B\stackrel{*,\ell}{\Rightarrow}a_{i}\ldots a_{k-1}$. Likewise, $\ell_2$ is the smallest integer for which $C\stackrel{*,\ell_2}{\Rightarrow}a_{k}\ldots a_{j-1}$. We consider all such productions in $P$ that will contribute tuples of the kind $(i,j,*)$ to $LIST(A)$ and from these only keep $(i,j,\ell)$ where $\ell$ is the least such integer. $\Box$

\section{Less than Cubic Time Parser}\label{lesscubic}
In this section we present an error correcting parser that runs in time
$O(T(n))$ where $n$ is the length of the input string and $T(n)$ is the time
needed to compute the tropical product of two $n\times n$ matrices. There are
two main ingredients in this parser, namely, the procedure given in
Section~\ref{ggen} for converting the given grammar into a covering grammar and
Valiant's reduction given in \cite{Val75} (and summarized in
Section~\ref{Valiant}).

As pointed out in Section~\ref{Valiant}, Valiant has presented an efficient
algorithm for computing the transitive closure of an upper triangular matrix.
The transitive closure is with respect to matrix multiplication defined in a
special way. Each element in a matrix will be a set of items. In standard
matrix multiplication we have two operators, namely, multiplication and
addition. In the case of special matrix multiplication, these operations are
replaced by $\odot$ and $\cup$ (called {\em union}). Valiant's algorithm works as long as these
operations satisfy the following properties: The outer operation (i.e., union)
is commutative and associative, the inner operation $\odot$ distributes over
union, $\emptyset$ is a zero with respect to $\odot$ and an identity with respect to union.

Valiant has shown that transitive closure under the above definition of matrix multiplication can be computed in $O(S(n))$ time, where $S(n)$ is the time needed for multiplying two matrices with the above special definition of matrix product.

In the context of error correcting parser we define the two operations as follows. Let ${\cal I}=w_1w_2\ldots w_n$ be the input string. Matrix elements are sets of pairs of the kind $(A,\ell)$ where $A$ is a nonterminal and $\ell$ is an integer. We initialize an $(n+1)\times(n+1)$ upper triangular matrix $a$  as:
\begin{equation}\label{eq1}
a_{i,i+1}=\{(A,\ell)|(A\stackrel{\ell}{\rightarrow}w_i)\in P'\},\ {\rm and}\ 
a_{i,j}=\emptyset,\ \ {\rm for}\ \ j\neq i+1.
\end{equation}

\begin{figure}[h]
\centering
\includegraphics[width=\textwidth]{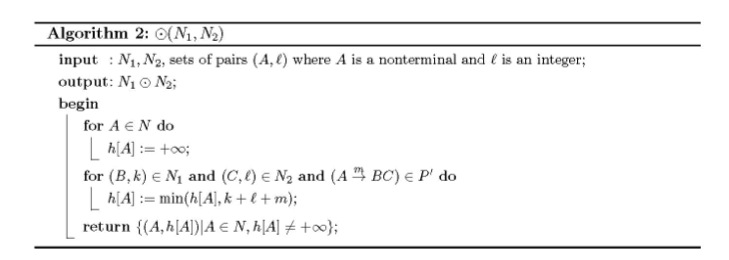}
\vspace*{-0.2in}
\caption{\label{fig2} Algorithm 2.}
\end{figure}

If $N_1$ and $N_2$ are sets of pairs of the kind $(A,\ell)$ then $N_1 \odot N_2$
is defined with the procedure in algorithm \ref{fig2}.

\vspace{0.2in}

\vspace{0.2in}

\commentOut{
\vspace{0.1in}

\vspace{0.1in}

\begin{quote}
$Q:=\emptyset$;\\
{\bf for} each $(B,k)\in N_1$ and each $(C,\ell)\in N_2$ {\bf do}\\
\hspace*{0.3in} {\bf if} $(A\stackrel{m}{\rightarrow}BC)\in P'$ {\bf then} add $(A,k+\ell+m)$ to $Q$;\\
 Sort the pairs in $Q$ in lexicographic nondecreasing order. In particular, $(A,k)\leq (B,\ell)$ iff $A\leq B$ and $k\leq \ell$. Let the sorted list of pairs in $Q$ be 
 $(A_1,l_1^1),(A_1,l_1^2),\ldots$, $(A_1,l_1^{n_1}),(A_2,l_2^1),\ldots,(A_2,l_2^{n_2}),\cdots,(A_q,l_1^1),\ldots,(A_q,l_q^{n_q})$.\\
 $N_1 \odot N_2=\{(A_1,l_1^1),(A_2,l_2^1),\ldots,(A_q,l_q^1)\}$.
\end{quote}

\vspace{0.1in}
}

\noindent If $N_1$ and $N_2$ are sets of pairs of the kind $(A,\ell)$, the union operation is defined as:
$$N_1\cup N_2=\{(A,k):(1)\ (A,k)\in N_1\ {\rm and}\ (A,k')\notin N_2\ {\rm for\ any}\ k'\ {\rm or}\ (2)\ (A,k)\in N_2\ {\rm and}$$
$$ (A,k')\notin N_1\ {\rm for\ any}\ k'\ {\rm or}\ (3)\ (A,k_1)\in N_1,(A,k_2)\in N_2\ {\rm and}\ k=\min(k_1,k_2)\}.$$

It is easy to see that the union operation is commutative and associative since if there are multiple pairs for the same nonterminal with different error counts, the union operation has the effect of keeping only one pair for each nonterminal with the least error count.

We can also verify that $\odot$ distributes over union. Let $N_1,N_2$, and
$N_3$ be any three sets of pairs of the type $(A,\ell)$. Let $Q=N_1 \odot
(N_2\cup N_3), R=(N_1 \odot N_2)\cup(N_1 \odot N_3),X=(N_1 \odot N_2)$, and
$Y=(N_1 \odot N_3)$.
If $(B,\ell)\in N_1,(C,k_1)\in N_2,(C,k_2)\in N_3,k=\min(k_1,k_2)$, and $(A\stackrel{m}{\rightarrow}BC)\in P'$, then $(A,k+\ell+m)\in Q$. Also, $(A,k_1+\ell+m)\in X$ and $(A,k_2+\ell+m)\in Y$. Thus, $(A,k+\ell+m)\in R$.

\begin{figure}[h]
\centering
\includegraphics[width=\textwidth]{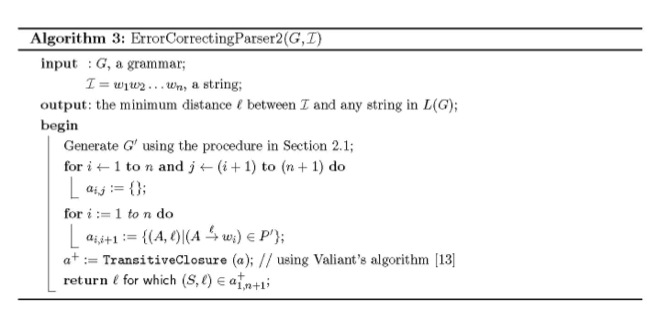}
\vspace*{-0.2in}
\caption{\label{fig3} Algorithm 3.}
\end{figure}

Put together, we get the following algorithm. Given a grammar $G$ and an input
string ${\cal I}$, generate the grammar $G'$ using the procedure in
Section~\ref{ggen}. Construct the matrix $a$ described in Equation~\ref{eq1}.
Compute the transitive closure $a^+$ of $a$ using Valiant's algorithm
\cite{Val75}. $(S,\ell)$ will occur in $a^+_{1,n+1}$ for some integer $\ell$. In
this case, the minimum distance between ${\cal I}$ and any string in $L(G)$ is
$\ell$. The pseudocode is given in algorithm \ref{fig3}.
 
\vspace{0.2in}
Note that, by definition,
$$a^+=a^{(1)}\cup a^{(2)}\cup\cdots$$
where $a^{(1)}=a$ and
$$a^{(i)}=\bigcup_{j=1}^{i-1}a^{(j)} \cdot a^{(i-j)}.$$

It is easy to see that $(A,\ell)$, where $A$ is a nonterminal and $\ell$ is an integer in the range $[0,n]$, will be in $a^{(k)}_{i,j}$ if and only if $A\derives[*,\ell] a_ia_{i+1}\cdots a_{j-1}$ such that $(j-i)=k$ and there is no $q<\ell$ such that $A\derives[*,q] a_ia_{i+1}\cdots a_{j-1}$. Also, $a^{(k)}_{i,j}=\emptyset$ if $(j-i)\neq k$. 

As a result, $(S,\ell)$ will be in $a^+_{1,n+1}$ if and only if  $S\derives[*,\ell] a_1a_{2}\cdots a_{n}$ and there is no $q<\ell$ such that $S\derives[*,q] a_1a_{2}\cdots a_{n}$.

We get the following

\begin{theorem}
Error correcting parsing can be done in $O(S(n))$ time where $S(n)$ is the time needed to multiply two matrices under the new definition of matrix product. $\Box$
\end{theorem}

It remains to be shown that $S(n)=O(T(n))$ where $T(n)$ is the time needed to compute the tropical product of two matrices.

Let $a$ and $b$ be two matrices where the matrix elements are sets of pairs of the kind $(A,\ell)$ where $A$ is a nonterminal and $\ell$ is an integer. Let $c=ab$ be the product of interest under the special definition of matrix product.

\begin{figure}[h]
\centering
\includegraphics[width=\textwidth]{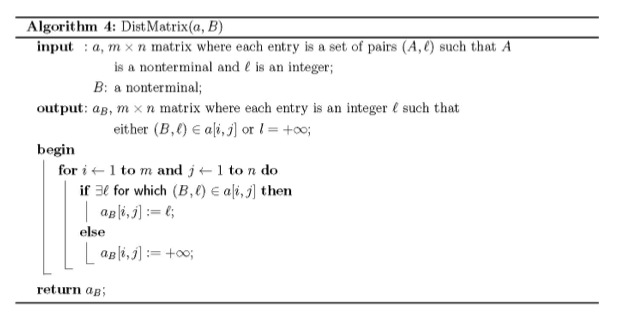}
\vspace*{-0.2in}
\caption{\label{fig4} Algorithm 4.}
\end{figure}

For each nonterminal $B$ in $G'$, we define a matrix $a_B$ and for each
nonterminal $C$ in $G'$, we define a matrix $b_C$. $a_B[i,j]=\ell$ if
$(B,\ell)\in a[i,j]$, for $1\leq i,j\leq n$. Likewise, $b_C[i,j]=\ell$ if
$(C,\ell)\in b[i,j]$, for $1\leq i,j\leq n$. The pseudocode of this
operation is given in algorithm \ref{fig4}. Compute the tropical product
$c_{BC}$ of $a_B$ and $b_C$ for every nonterminal $B$ and every nonterminal $C$.

\begin{figure}[h]
\centering
\includegraphics[width=\textwidth]{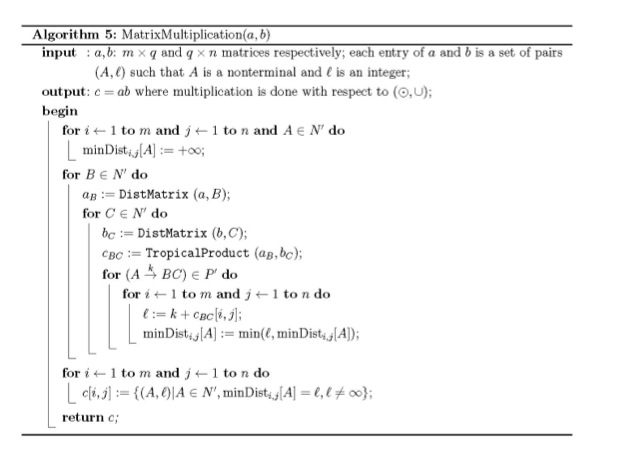}
\vspace*{-0.2in}
\caption{\label{fig5} Algorithm 5.}
\end{figure}

For every production $A\stackrel{k}{\rightarrow} BC$ in $P'$ do the following:
for every $1\leq i,j\leq n$, if $c_{BC}[i,j]=\ell$ then add $(A,k+\ell)$ to
$c[i,j]$, keeping only the smallest distance if $A$ is already present in
$c[i,j]$.
\commentOut{
Now, go through each element in $c$ and eliminate duplicates. Specifically, let $Q$ be any element of $c$. Sort the pairs in $Q$ in lexicographic order. Let the sorted list of pairs in $Q$ be 
 $(A_1,l_1^1),(A_1,l_1^2),\ldots$, $(A_1,l_1^{n_1}),(A_2,l_2^1),\ldots,(A_2,l_2^{n_2}),\cdots,(A_q,l_1^1),\ldots,(A_q,l_q^{n_q})$. Replace $Q$ with
 $\{(A_1,l_1^1),(A_2,l_2^1),\ldots,(A_q,l_q^1)\}$. 
 }
 The pseudocode is given in algorithm \ref{fig5}.
 
 Clearly, the time spent in computing $ab$ (under the new special definition of matrix product) is $O(T(n)+n^2)$ assuming that the size of the grammar is $O(1)$. 
 
 Put together we get the following theorem.
 
 \begin{theorem}
 Error correcting parsing can be done in $O(T(n))$ time where $T(n)$ is the time needed to compute the tropical product of two matrices. $\Box$
 \end{theorem}
 
 Using the currently best known algorithm for tropical products \cite{Chan10},
 we get the following theorem.
 
  \begin{theorem}
 Error correcting parsing can be done in $O\left(\frac{n^3(\log\log n)^3}{(\log n)^2}\right)$ time. $\Box$
 \end{theorem}
 
 Furthermore, consider the case of error correcting parsing where we know a
 priori that there exists a string ${\cal I}'$ in $L(G)$ such that the distance
 between $\cal I$ and ${\cal I}'$ is upper bounded by $m$. We can solve this
 version of the language edit distance problem using the tropical matrix product
 algorithm of Zwick~\cite{Zwick98}. This algorithm multiplies two $n\times n$
 integer matrices in $O(Mn^\omega)$ time if the matrix elements are in the
 range $[-M,M]$  \cite{Zwick98}. Here $O(n^\omega)$ is the time taken to
 multiply two $n\times n$ real matrices. Recall that when we reduce the matrix
 multiplication under $\odot$ and $\cup$ to tropical matrix multiplication, we have to compute the tropical product $c_{BC}$ of $a_B$ and $b_C$ for every nonterminal $B$ and every nonterminal $C$. Elements of $a_B$ and $b_C$ are integers in the range $[0,n]$. Note that even if all the elements of $c_{BC}$ are $\leq m$, some of the elements of $a_B$ and $b_C$ could be larger than $m$. Before using the algorithm of \cite{Zwick98} we have to ensure that all the elements of $a_B$ and $b_C$ are less than $M$ (where $M$ is some function of $m$). This can be done as follows.
 Before invoking the algorithm of \cite{Zwick98}, we replace every element of $a_B$ and $b_C$ by $m+1$ if the element is $>m$. $m+1$ is 'infinity' as far as this multiplication is concerned. By doing this replacement, we are not affecting the final result of the algorithm and at the same time, we are making sure that the elements of $a_B$ and $b_C$ are $\leq M=(m+1)$.
 
 As a result, we get the  following theorem.
 
 \begin{theorem}
 Error correcting parsing can be done in $O\left(mn^\omega\right)$ time where
 $m$ is an upper bound on the edit distance between the input string $\cal I$ and some string ${\cal I}'$ in $L(G)$, $G$ being the input CFG.
 $O(n^\omega)$ is the
 time it takes to multiply two $n \times n$
 matrices. $\Box$
 \end{theorem}
 
 As a corollary to the above theorem we can also get the following theorem.
 
 \begin{theorem}
 There exists an $\frac{n}{m}$-approximation algorithm for the language edit distance problem that has a run time of $O(mn^\omega)$, where $O(n^\omega)$ is the time taken to multiply two $n\times n$ matrices. 
 \end{theorem}
 
 \noindent{\bf Proof:} Here again, we replace every element of $a_B$ and $b_C$ by $m+1$ if the element is $>m$. In this case the elements of $c_{BC}$ will be $\leq(2m+2)$. We replace any element in $c_{BC}$ that is larger than $m$ with $(m+1)$. In general whenever we generate or operate on a matrix, we will ensure that the elements are $\leq (m+1)$. If $S\stackrel{\ell}{\Rightarrow}{\cal I}$ for some $\ell\leq m$, then the final answer output will be exact. If $\ell>m$, then the algorithm will always output $n$. Thus the theorem follows. $\Box$
 
 \section{Retrieving ${\cal I}'$}
 In all the algorithms presented above, we have focused on computing the minimum edit distance between the input string $\cal I$ and any string ${\cal I}'$ in $L(G)$. In this section we address the problem of finding ${\cal I}'$. We show that ${\cal I}'$ can be found in $O(n^2)$ time, where $n=|{\cal I}|$. Let $S\stackrel{*,\ell}{\Rightarrow}{\cal I}$ such that there is no $k<\ell$ such that $S\stackrel{*,k}{\Rightarrow}{\cal I}$. Let ${\cal I}=a_1a_2\cdots a_n$.
 
\begin{figure}[h]
\centering
\includegraphics[width=\textwidth]{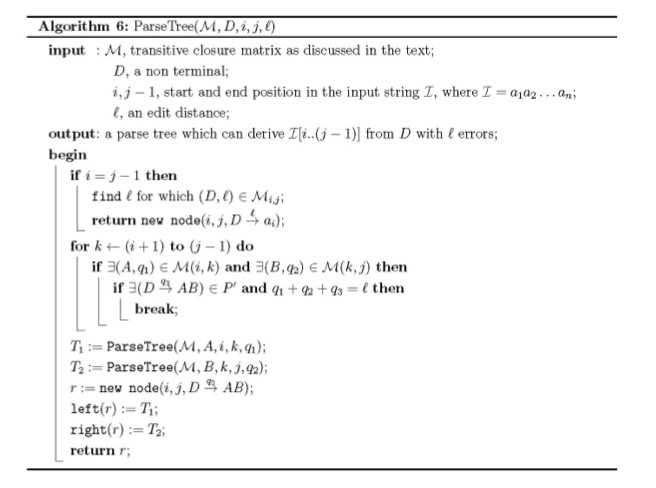}
\vspace*{-0.2in}
\caption{\label{fig6} Algorithm 6.}
\end{figure}

 Realize that in the algorithms given in Section~\ref{simple} and Section~\ref{lesscubic} we compute, for every $i$ and $j$ (with $j>i$), all the nonterminals $A$ such that $A$ spans $a_{i}a_{i+1}\ldots a_{j-1}$ and we also determine the least $k$ such that $A\stackrel{*,k}{\Rightarrow}a_{i}a_{i+1}\ldots a_{j-1}$. In this case, there will be an entry for $A$ in the matrix $\cal M$. Specifically, $(A,k)$ will be in ${\cal M}(i,j)$. We can utilize this information to identify an ${\cal I}'$ such that the edit distance between $\cal I$ and ${\cal I}'$ is equal to $\ell$. Note that we can deduce ${\cal I}'$ if we know the sequence of productions used to derive ${\cal I}'$.
The pseudocode is given in algorithm \ref{fig6}. We will invoke the
algorithm as $ParseTree({\cal M}, S, 1, n+1, \ell)$.

\commentOut{
 Similar to the algorithm in Section~\ref{simple}, we can develop an algorithm
 with $\leq n$ stages. In the first stage we consider every production
 $S\stackrel{q_1}\rightarrow AB$ in $P'$ and do the following.
 
 \begin{quote}
 {\bf for} $i:=1$ {\bf to} $n-1$ {\bf do}
 
 \hspace{0.3in} If $(A,q_2)\in {\cal M}(1,i)$ and $(B,q_3)\in {\cal M}(i,n)$ and $q_1+q_2+q_3=\ell$, use $S\stackrel{q_1}\rightarrow AB$ 
 
 \hspace{0.3in} as the first production needed to deduce ${\cal I}'$. 
 
 \hspace{0.3in} Call this value of $i$ as $j$ and quit;
 \end{quote}
 }
 
Algorithm 6 finds the first production in $O(n)$ time. Having found the first production, we can proceed in a similar manner to find the other productions needed to derive ${\cal I}'$. In the second stage we have to find a production that can be used to derive $a_1a_2\ldots a_j$ from $A$ and another production that can be used to derive $a_{j+1}a_{j+2}\ldots a_n$ from $B$. Note that the span length of $A$ plus the span length of $B$ is $n$ and hence both the productions can be found in a total of $O(n)$ time. 

We can think of a tree $\cal T$ where $S$ is the root and $S$ has two children $A$ and $B$. If $A\rightarrow CD$ is the first production that can be used to derive  $a_1a_2\ldots a_j$ from $A$ and $B\rightarrow EF$ is the first production that can be used to derive  $a_{j+1}a_{j+2}\ldots a_n$ from $B$, then $A$ will have two children $C$ and $D$ and $B$ will have two children $E$ and $F$. 

The rest of the tree is constructed in the same way.  Clearly, the total span length of all the nonterminals in any level of the tree is $\leq n$ and hence the time spent at each level is $O(n)$. Also, there can be at most $n$ levels. As a result, we get the following theorem.

\begin{theorem}
We can identify ${\cal I}'$ in $O(n^2)$ time. $\Box$
\end{theorem}

\section{Conclusions}
In this paper we have presented an error correcting parser for general context free languages. This algorithm takes less than cubic time, improving the 1972 algorithm of Aho and Peterson that has remained the best until now. We have also shown that if $M$ is an upper bound on the edit distance between the input string $\cal I$ and some string of $L(G)$,  then we can solve the parsing problem in $O(Mn^\omega)$ time, where $O(n^\omega)$ is the time it takes to multiply two $n\times n$ matrices. As a corollary, we have presented an $\frac{n}{M}$-approximation algorithm for the general context free language edit distance problem that runs in time $O(Mn^\omega)$.

\section*{Acknowledgements}
This work has been supported in part by the following grant: NIH R01LM010101. The first author thanks Barna Saha for the introduction of this problem and Alex Russell for providing pointers to tropical matrix multiplication.

\end{document}